\documentclass[pre,epsfig,twocolumn,showpacs,preprintnumbers,amssymb]{revtex4}
\usepackage{bm,graphicx,amsmath}

\begin{document}

\preprint{published in Eur.\ Phys.\ J.\ E {\bf 18}, 123-131 (2005)}

\title{Photon Channelling in Foams}

\author{Michael Schmiedeberg $^{1}$}
\author{MirFaez Miri $^{2}$}
\author{Holger Stark $^1$}
\affiliation{$^1$ Universit\"at Konstanz, Fachbereich Physik, D-78457
             Konstanz, Germany\\
             $^2$ Institute for Advanced Studies in Basic Sciences, 
             Zanjan, 45195-159, Iran}


\begin{abstract}
Experiments by Gittings, Bandyopadhyay, and Durian 
[Europhys.\ Lett.\ \textbf{65}, 414 (2004)]
demonstrate that light possesses a higher probability to propagate in the 
liquid phase of a foam due to total reflection. The authors term this 
observation photon channelling which we investigate in this
article theoretically.
We first derive a central relation in the work of Gitting {\em et al.}
without any free parameters. It links the photon's path-length fraction 
$f$ in the liquid phase to the liquid fraction $\varepsilon$.
We then construct
two-dimensional Voronoi foams, replace the cell edges by channels
to represent the liquid films and simulate photon paths according to
the laws of ray optics using transmission and reflection coefficients
from Fresnel's formulas. In an exact honeycomb foam, the photons show
superdiffusive behavior. It becomes diffusive as soon as disorder is 
introduced into the foams. The dependence of the diffusion constant on 
channel width and refractive index is explained by a one-dimensional 
random-walk model. It contains a photon channelling state that is crucial 
for the understanding of the numerical results. At the end, we shortly
comment on the observation that photon channelling only occurs in a 
finite range of $\varepsilon$.
\end{abstract}



\maketitle


\section{Introduction}\label{intro}

Aqueous foams consist of gas bubbles separated by liquid films\
\cite{Weaire1999}. In dry 
foams, these bubbles are deformed to polyhedra. Always three films meet
in the so-called Plateau borders which form a network of liquid
channels throughout the foams. Their opaque appearance identifies
them as multiply scattering media. Moreover, careful light-scattering 
experiments show that light transport has reached its
diffusive limit in foams\ \cite{Durianold,Ern,Hoehler,Gopal99,durApp} 
which means that photons perform 
a random walk. Diffusion of light is well established
in colloidal systems\ \cite{DWS}, where light is scattered from
particles, or in nematic liquid crystals\ \cite{DWSnematic}, 
where scattering occurs from fluctuations in the local optical axis. 
Diffusing-wave\ \cite{DWS} and 
diffuse-transmission spectroscopy\ \cite{Kaplan94} provide non-invasive 
probes of structure and dynamics in bulk samples. While these methods have 
already successfully been applied to monitor the internal dynamics
of foams\ \cite{Durianold,Ern,Hoehler,Gopal99},
no clear understanding has emerged so far about the basic mechanism
underlying the random walk of photons in these cellular structures. 
Estimates for the scattering from Plateau borders do not seem to be
in accordance with experiments\ \cite{durApp,Gittings2004}. 
Scattering from vertices formed by four meeting Plateau borders would
be another possibility\ \cite{Skipetrov02}.
On the other hand, since gas bubbles in foams are much larger than the 
wavelength of light, one can use the laws of ray optics to follow the
photons on their random walk as they are reflected by the liquid
films\ \cite{Miri03,Miri04,Miri05,Miri05a}.

The work presented in this paper was initiated by a very inspiring 
publication of Gittings, Bandyopadhyay, and Durian published under the 
same title\ \cite{Gittings2004}. The authors argued that total
reflection of light at the liquid-gas interface in a foam should
increase the probability of photons to propagate in the liquid phase,
made up by thin films and Plateau borders, and they termed this
effect photon channelling. By measuring the absorption of light in
foams with strongly absorbing liquid, they determined the fraction $f$
of a photon's path that lies in the liquid. Equal distribution of the
photons throughout the foam would give $f=\varepsilon$, where $\varepsilon$
is the liquid fraction in the foam. However, the authors found that
in a certain range of $\varepsilon$
the path-length fraction $f$ fulfills $f > \varepsilon$. This not only 
confirms photon channelling but it also gives clear evidence that the 
basic mechanism for light transport in foams depends on the liquid
volume fraction. 

Motivated by these findings, we perform a thorough theoretical study
of photon channelling in this article based on analytical arguments
and numerical simulations using ray optics.
Gittings {\em et al.} derived a relation between $f$ and $\varepsilon$
\cite{Gittings2004}.
It contains the ratio of averaged transmission probabilities which
they determined both by experiments and simulations. In Sec.\ 
\ref{channel.analytic}, we demonstrate how this ratio follows from
general arguments and are therefore able to justify the relation 
$f(\varepsilon)$ on completely analytical grounds. Furthermore,
Gittings {\em et al.} stated that the effect of photon channelling
would prevent the photons from performing a ``truely random walk''\
\cite{Gittings2004}. Indeed, our numerical investigation in 
Sec.\ \ref{superdiff} shows that transport of light in a perfect
honeycomb foam is superdiffusive. The effect is most pronounced when the 
liquid-gas interface is totally reflecting\ \cite{Schmiedeberg05b}.
However, the transport becomes diffusive as soon as the foam contains disorder.
This demonstrates that photon channelling is compatible with the
concept of random walk. To complete our numerical study of photon
channelling, we have developed a one-dimensional random walk model.
It contains a photon-channelling state and can therefore explain the
main features of our numerical investigation.

The paper is organized as follows. In Sec.\ \ref{model}, we present 
two-dimensional model foams for our photon-channelling study and
explain how we simulate the photons' random walk based on the laws of 
ray optics.
In Sec.\ \ref{channel}, characteristic features of photon channelling
are first explained analytically and then compared with simulation results. 
The transport properties of light in our model foams are described in 
Sec.\ \ref{photondiff}. Numerical results on superdiffusion in
perfect honeycomb structures and diffusion in disordered foams are
presented. Finally, we introduce the one-dimensional random-walk model
for photon channelling. Section\ \ref{conclus} summarizes and
discusses our results.

\section{Model System and Method of Simulation} \label{model}

To introduce a simple model for a two-dimensional foam, we start from
a Voronoi tessellation of the plane\ \cite{riv,okabe}. It is generated from
a distribution of seed points in a simulation box, for which Voronoi 
polygons are constructed in complete analogy to the Wigner-Seitz cells 
of periodically arranged lattice sites. We begin with a triangular 
lattice of seed points that gives a regular honeycomb structure whose edges
possess the length $l_{0}$. Then we systematically introduce disorder by 
shifting the seed points along a randomly chosen displacement vector 
whose magnitude is equally distributed in the 
intervall $[0,\delta r]$. Examples of Voronoi foams are presented, e.g.,
in Ref.\ \cite{Miri04}. 
Note, while the structure and evolution of real foams are constrained by
Plateau's rules and Laplace's law\ \cite{Weaire1999},
simple honeycomb, Voronoi, and three-dimensional Kelvin structures were
used for a first study of the physical properties of foams\
\cite{Gibson, HVK}.

\begin{figure}
\includegraphics[width=0.6\columnwidth]{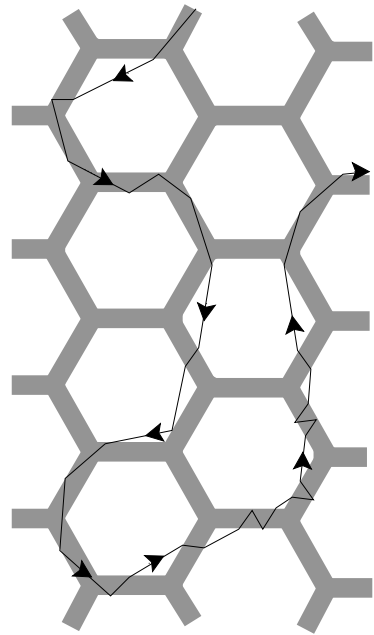}
\caption{Regular honeycomb foam ($\delta r = 0$) and a photon's random walk 
in this structure.}
\label{fig:channel}
\end{figure}

All our Voronoi tessellations are produced by the 
software {\it Triangle}\ \cite{triangle}. They contain approximately
$15000$ cells which corresponds to a quadratic simulation box with edge 
length 200 $l_{0}$. 
Now, on each edge of the Voronoi tessellation, we place a channel of
width $d$ that represents a liquid film. Our final model foam for 
$\delta r = 0$ is 
indicated in Fig.\ \ref{fig:channel}. For this regular honeycomb foam, we 
calculate the fraction of the plane filled by the liquid phase as
\begin{equation}
\varepsilon = \frac{2}{\sqrt{3}}\frac{d}{l_{0}} \left( 1-\frac{1}{2\sqrt{3}}
\frac{d}{l_{0}} \right) \enspace.
\label{eqn:eps}
\end{equation}
In the following, we only investigate foams with modest disorder 
quantified by $\delta r \le 0.3 l_{0}$. This avoids the unphysical situation 
that four instead of three channels or films meet when we construct
our model foams. All cells still have six edges. Furthermore, to simulate 
photon diffusion in the model foams, periodic boundary conditions are used.

We let photons perform a random walk in our model foams by applying the
rules of geometrical optics (see Fig.\ \ref{fig:channel}). 
The photons move straight in the liquid
or gaseous phase with respective velocities $c/n_{l}$ or $c/n_{g}$
where $c$ is the vacuum speed of light and $n_{l},n_{g}$ are refractive
indices. When the photons hit a liquid-gas interface, they are reflected 
with a probability, also called reflectance, given by Fresnel's formulas. 
The respective reflectances for electric polarizations parallel or 
perpendicular to the plane of incidence are
\begin{equation}
\label{eqn:fresnel}
r_{\|} = \left( 
  \frac{\mathrm{tan}(\beta_{l}-\beta_{g})}{\mathrm{tan}(\beta_{l}+\beta_{g})}
                              \right)^{2} \enspace \mathrm{and} \enspace
r_{\perp} = \left( 
  \frac{\mathrm{sin}(\beta_{l}-\beta_{g})}{\mathrm{sin}(\beta_{l}+\beta_{g})}
                              \right)^{2} \enspace,
\end{equation}
where $\beta_{l}$ and $\beta_{g}$ denote the respective angles of a 
light ray in the liquid or gaseous phase measured relative to the normal 
on the interface. They are connected by Snellius' law:
$\mathrm{sin} \beta_{l}/ \mathrm{sin} \beta_{g} = n_{g}/n_{l}$.
Note that Fresnel's formulas are symmetric in $\beta_{l}$ and 
$\beta_{g}$. So the reflectance is the same whether the light ray is
hitting the interface coming from the liquid or the gaseous phase. However, 
there is an important difference: light rays in the optically denser
medium ($n_{l} > n_{g}$) experience total reflection for an incident 
angle $\beta_{l} > \beta^{\ast} = \mathrm{arcsin}(n_{g}/n_{l})$. We 
stress this point here because it is central for the occurence of photon
channelling.

Typically, we launch 10000 photons at one vertex of the underlying
Voronoi tesselation in an angular range of $60^{\circ}$ and let them run 
during a time $t=10^{5}l_{0}/c$. At several times, we calculate the 
mean square displacement $\sigma^{2}$ from the photon cloud and plot it 
as a function of $t$. Diffusion constants are then determined from a fit 
to $\sigma^{2} = 4 D t$.

\section{Photon Channelling} \label{channel}

At the core of the observation of photon channelling is the fact that
photons can totally be reflected when they hit the liquid-gas interface 
from the liquid side. Accordingly, a photon spends more time in the 
liquid phase or, more accurately, the fraction $f$ of the photon's path 
that lies in the liquid phase is not equal to the liquid volume fraction 
$\varepsilon$. Gittings {\em et al} measured $f$
with the help of a strongly absorbing liquid and then made a prediction
for the two averaged transmission probabilities for photons going from the 
liquid to the gaseous phase ($T_{l\rightarrow g}$) or the reverse 
direction ($T_{g\rightarrow l}$). Only in the range $0.04 < \epsilon < 0.2$,
they found that the ratio $T_{l\rightarrow g}/T_{g\rightarrow l}$
deviates from one, indicating photon channelling. Furthermore they gave its
approximate value as 
$T_{l\rightarrow g}/T_{g\rightarrow l} \approx n_{g}/n_{l}$ and also
confirmed it by simulations.
In the following, we derive this central relation for 
$T_{l\rightarrow g}/T_{g\rightarrow l}$ analytically, connect it to $f$ 
and compare it with our simulations.

\subsection{Analytic results} \label{channel.analytic}

We start with defining the averaged transmission probability 
$T_{l\rightarrow g}$ by
\begin{equation}
T_{l\rightarrow g}=\int_{0}^{\pi/2} p_{l}(\beta_{l}) t(\beta_{l},\beta_{g})
d\beta_{l} \enspace,
\label{3}
\end{equation}
where the angle-dependent transmission coefficient 
$t(\beta_{l},\beta_{g}) = 1-r(\beta_{l},\beta_{g})$ is connected to the
reflectance $r$ of Fresnel's formulas introduced in Eq.\ (\ref{eqn:fresnel}). 
Note that only the angular range of $\beta_{l} < \beta^{\ast}$ contributes to
the integral since $t(\beta_{l},\beta_{g})$ is $0$ for 
$\beta_{l} > \beta^{\ast}$ due to total reflection. 
The factor $p_{l}(\beta_{l})$ is the normalised probability
of a photon to reach the interface from the liquid side with an 
incident angle $\beta_{l}$. We write it as
\begin{equation}
p_{l}(\beta_{l}) = P_{l}(\beta_{l}) / 
\int_{0}^{\pi/2}P_{l}(\beta_{l})d\beta_{l} \enspace.
\label{4}
\end{equation}
Accordingly, we define the transmission probability $T_{g \rightarrow l}$ by
\begin{equation}
T_{g \rightarrow l} = \int_{0}^{\beta^{\ast}} p_{g}(\beta_{l}) 
t(\beta_{l},\beta_{g}) d\beta_{l} \enspace,
\label{5}
\end{equation}
where we used the symmetry of Fresnel's formulas with respect to 
$\beta_{l} \leftrightarrow \beta_{g}$. We still take the angle $\beta_{l}$
for the averaging, therefore the upper limit of the integral is just 
$\beta^{\ast}$ since photons entering the liquid cannot
have a $\beta_{l}$ larger than $\beta^{\ast}$. For $\beta_{l} < \beta^{\ast}$
it makes sense to assume that photons coming from the gaseous into 
the liquid phase possess the same angular distribution as photons in the 
liquid, i.e., $P_{g}(\beta_{l}) \propto P_{l}(\beta_{l})$. So we write
\begin{eqnarray}
p_{g}(\beta_{l}) & = & P_{g}(\beta_{l}) / 
  \int_{0}^{\beta^{*}}P_{g}(\beta_{l})d\beta_{l} \nonumber \\
                 & = & P_{l}(\beta_{l}) / 
  \int_{0}^{\beta^{*}}P_{l}(\beta_{l})d\beta_{l} \enspace.
\label{6}
\end{eqnarray}
Note again the different upper limit in the integral of the normalization 
factor compared to Eq.\ (\ref{4}). Since $t(\beta_{l},\beta_{g})$ is $0$ 
for $\beta_{l} > \beta^{\ast}$, the transmission probabilities
of Eqs. (\ref{3}) and (\ref{5}) appear the same however they differ by
the normalization factors of $p_{l}$ and $p_{g}$ in 
Eqs.\ (\ref{4}) and (\ref{6}) as just mentioned.
Therefore the ratio of the averaged transmission probabilities is
\begin{equation}
\frac{T_{l\rightarrow g}}{T_{g\rightarrow l}} = 
\frac{\int_{0}^{\beta^{*}}P_{l}(\beta_{l}) d\beta_{l}}{
          \int_{0}^{\pi/2}P_{l}(\beta_{l})d\beta_{l}} \enspace.
\label{7}
\end{equation}

We now assume that 
all directions of the photons are equally distributed in the liquid and 
therefore obtain for the angular distribution of photons reaching the
interface, $P_{l}(\beta_{l})=\cos\beta_{l}$.
The term $\cos\beta_{l}$ means that the number of photons
hitting a given surface area decreases with increasing $\beta_{l}$,
i.e., photons approaching the interface on a ``shallow'' path 
($\beta_{l} \rightarrow \pi/2$) have a vanishing probability to hit
a given surface area.
With $\beta^{\ast} = \mathrm{arcsin}(n_{g}/n_{l})$, the ratio of the mean 
transmission probabilities becomes immediately
\begin{equation}
\frac{T_{l\rightarrow g}}{T_{g\rightarrow l}}=\frac{\int_{0}^{\beta^{*}}
\cos\beta_{l} d\beta_{l}}{\int_{0}^{\pi/2}\cos\beta_{l}d\beta_{l}} = 
\frac{n_{g}}{n_{l}} \enspace.
\label{8}
\end{equation}
This is the result of Gittings {\em et al.}\ \cite{Gittings2004} measured
in the range $0.04 < \epsilon < 0.2$ for the liquid fraction.
In our simulations we realize $P_{l}(\beta_{l}) = \cos\beta_{l}$ by starting 
the photons in the liquid with a uniform angular distribution. In
experiments, photons always enter from the gaseous phase. A fast 
randomization of the angular distribution in the liquid phase is then 
achieved by disorder in the foam and by the strong curvature of the 
interface at the Plateau borders. Such a randomization due to curvature 
occurs, e.g., in the Sinai billard\ \cite{Sinai}.

We now derive how the fraction $f$ of a photon's path that lies in the 
liquid phase depends on the liquid volume fraction $\varepsilon$.
We assume a stationnary photon distribution in the foam. 
Then the photon current densities $j_{l\rightarrow g}$ and 
$j_{g\rightarrow l}$  across an interface have to be equal: 
$j_{l\rightarrow g} = j_{g\rightarrow l}$. With 
$j_{l\rightarrow g} = (c/n_{l}) \varrho_{l} T_{l\rightarrow g}$
and $j_{g\rightarrow l} = (c /n_{g}) \varrho_{g} T_{g\rightarrow l}$,
where $\varrho_{i}$ is the number density of the photons in phase $i$, and
using relation\ (\ref{8}), we find immediately
\begin{equation}
\frac{\varrho_{l}}{\varrho_{g}} = \frac{n_{l}^{2}}{n_{g}^{2}} \enspace.
\label{9}
\end{equation}
The ratio of the average times $\tau_{l}$ and $\tau_{g}$ a photon spends in
the liquid or gas equals the ratio of the average number of photons
in these phases,
\begin{equation}
\frac{\tau_{l}}{\tau_{g}} = 
\frac{\varrho_{l} V_{l}}{\varrho_{g} V_{g}} \enspace,
\label{10}
\end{equation}
where $V_{l}$ and $V_{g}$ denote the respective volumes. To motivate this
equation, we consider the average photon number $\varrho_{i} V_{i}$ as
a time average over the paths of many photons so that 
$\tau_{i} \propto \varrho_{i} V_{i}$.
In experiments, the average path-length fraction 
$f = s_{l} /(s_{l}+s_{g})$ is measured. Here $s_{i}$ is the path length
in phase $i$. With $s_{i} = (c/n_{i}) t_{i}$ and Eqs.\ (\ref{9}), (\ref{10}),
we finally obtain
\begin{equation}
f^{-1} = \frac{s_{g}+s_{l}}{s_{l}} = 1 + \frac{n_{g}}{n_{l}}
\left(\frac{1}{\varepsilon} - 1 \right) 
\label{11}
\end{equation}
where $\varepsilon = V_{l} /(V_{l}+V_{g})$.
With the actual values of $n_{g} =1$ and $n_{l} = 1.33$, i.e., 
$n_{g}/n_{l} = 0.75$, we obtain the semi-empirical relation for $f^{-1}$ that
Gittings {\em et al.} report in Ref.\ \cite{Gittings2004}.

If we do not specify the value of $T_{l\rightarrow g}/ T_{g\rightarrow l}$
in the derivation of Eq.\ (\ref{11}), we obtain the relation
\begin{equation}
\frac{T_{l\rightarrow g}}{T_{g\rightarrow l}} = 
\frac{\varepsilon (1-f)}{f(1-\varepsilon)}
\label{12}
\end{equation}
that was already given by Gittings {\em et al.}\ \cite{Gittings2004}. It
links $T_{l\rightarrow g}/ T_{g\rightarrow l}$ to the measurable
quantity $f$.

\subsection{Numeric results} \label{channel.numeric}

\begin{figure}
\includegraphics[width=0.49\columnwidth]{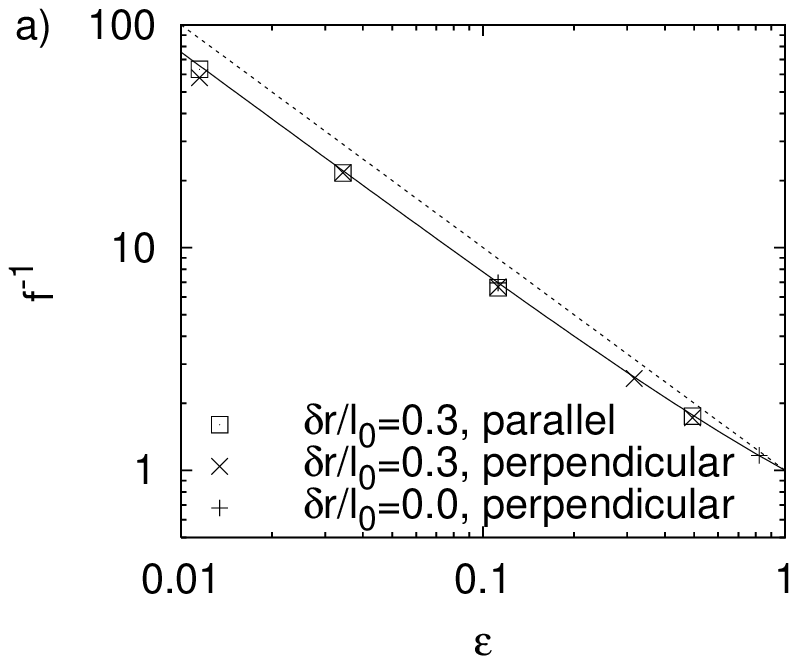}
\includegraphics[width=0.49\columnwidth]{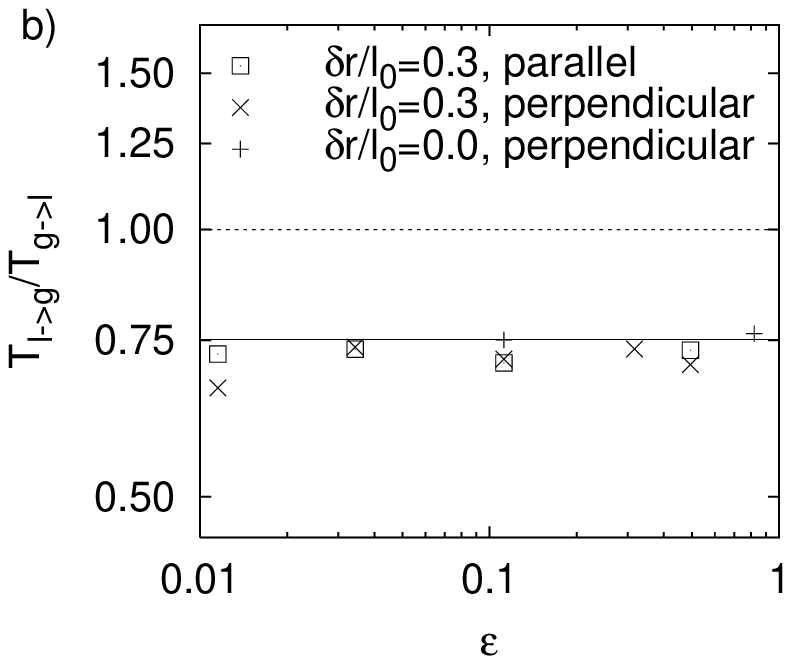}
\caption{a) Inverse of the path-length fraction $f$ as a function of
liquid fraction $\varepsilon$ for a disordered model foam 
($\delta r/l_{0}=0.3$) and a perfect honeycomb structure
($\delta r/l_{0}=0.0$). Photons with 
parallel or perpendicular polarization states are chosen. The full line 
corresponds to Eq.\ (\ref{11}) and the dashed line to 
$f^{-1}=\varepsilon^{-1}$. Further parameters are $n_{g}=1$ and 
$n_{l} = 1.33$. b) The ratio of the average transmission probabilities 
$T_{l\rightarrow g}/ T_{g\rightarrow l}$ as a function of $\varepsilon$. 
Otherwise, same description as in a).}
\label{fig:fraction}
\end{figure}

We performed simulations to confirm the analytic results of the
previous section always choosing $n_{g}=1$. 
For $n_{l} = 1.33$, we plot in Fig.\ \ref{fig:fraction}a)
the inverse of the path-length fraction $f$ as a function of
the liquid fraction $\varepsilon$ which was calculated via 
Eq.\ (\ref{eqn:eps}) from the channel width $d/l_{0}$. 
A perfectly ordered model foam
and one with disorder $\delta r = 0.3l_{0}$ were chosen. The photons
were either in the parallel or perpendicular polarization state. The 
numerical results represented by the symbols agree very well with the 
analytic prediction of Eq.\ (\ref{11}) (full line). Therefore, the diagram 
and also further simulations clearly show that the validity of 
Eq.\ (\ref{11}) is independent of disorder in the model foam and the 
polarization state. The dashed line corresponds to $f=\varepsilon$.
According to Eq.\ (\ref{11}) this means $n_{g}=n_{l}$, i.e., when
there is no optical contrast between the cells and the channels.
Differently speaking, photon channelling should always occur in this
model for $n_{g} \ne n_{l}$. In Fig.\ \ref{fig:fraction}b), we plot 
$T_{l\rightarrow g}/ T_{g\rightarrow l}$, calculated from Eq. (\ref{12}),
as a function of $\varepsilon$ using the results of 
Fig.\ \ref{fig:fraction}a). Again the different numerical data points
agree with our theory culminating in Eq.\ (\ref{8}) and therefore
justify the assumptions made during the derivation of Eq.\ (\ref{8}).
Finally, we also checked that the effect of photon channelling becomes 
stronger for increasing $n_{l}$ as expected.

\section{Photon Diffusion} \label{photondiff}

\subsection{Superdiffusion in Honeycomb Foams} \label{superdiff}

\begin{figure}
\includegraphics[width=0.97\columnwidth]{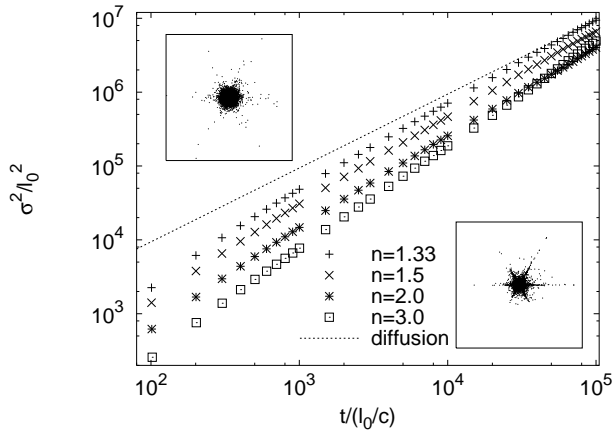}
\caption{Exact honeycomb foam: Reduced mean-square displacement 
$\sigma^{2}/l_{0}^{2}$ of the photon cloud as a function of reduced time 
$t/(c/l_{0})$ for different refractive indices $n_{l}$ of the liquid 
phase and for $d/l_{0}=1$. Perpendicular polarization is chosen. The dashed 
line indicates pure diffusion. Insets: Photon clouds at $t=10^{5} l_{0}/c$ 
for $n_{l} = 1.33$ (top left) and $3.0$ (bottom right).}
\label{fig:super1}
\end{figure}

Here we investigate the behavior of photon propagation in the exact
honeycomb foam of Fig.\ \ref{fig:channel}. In Fig.\ \ref{fig:super1} 
we plot the mean-square displacement $\sigma^{2}$
of the photon cloud (in units of $l_{0}^{2}$) as a function of time $t$ 
(in units of $l_{0}/c$) for different refractive indices $n_{l}$ of 
the liquid phase. The channel-width-to-length ratio is $d/l_{0}=1$.
As a reference, the dashed line indicates pure 
diffusive behavior with $\sigma^{2} \propto t$. Clearly, in the exact 
honeycomb foam the
spreading of the photons is superdiffusive. The effect is strong for
short times with an exponent $m$ in $\sigma^{2} \propto t^{m}$ between
1.5 and 1.6. It becomes weaker for large times, e.g., $m=1.1$ for 
$n_{l}=1.33$. In the framework of L\'evy walks, superdiffusion is
associated with a distribution of step lengths whose first or second moment
(mean value and variance) do not exist \cite{Levywalk}.
When the boundaries of the liquid channels are completely reflecting,
photon paths occur that correspond to effective steps of infinite length
along the directions of the channels\ \cite{Schmiedeberg05b}. In such a 
system, superdiffusion is most pronounced also for large times as we shall 
demonstrate in a forthcoming publication\ \cite{Schmiedeberg05b}.
It becomes weaker by letting the photons enter the gaseous phase 
since this is a mean to reduce the number of very long or infinite photon 
steps. Note that in Fig.\ \ref{fig:super1} superdiffusion is less
pronounced for smaller $n_{l}$, i.e., when the reflectance at the
liquid-gas interface decreases. Another mean to cut long steps would 
be rounding off the sharp edges in the channel system. 
Finally, as demonstrated in the next section, introducing disorder in the 
model foam leads to conventional photon diffusion as observed in experiments.

\begin{figure}
\includegraphics[width=0.97\columnwidth]{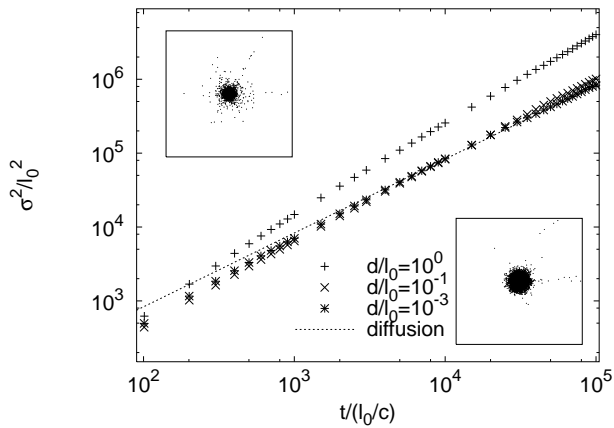}
\caption{Exact honeycomb foam: Reduced mean-square displacement 
$\sigma^{2}/l_{0}^{2}$ of the photon cloud as a function of reduced time 
$t/(c/l_{0})$ for different ratios $d/l_{0}$. The refractive index is $n=2$
and perpendicular polarization is chosen. The dashed line indicates 
pure diffusion. Insets: Photon clouds at $t=10^{5} l_{0}/c$ for 
$d/l_{0} = 1$ (top left) and $10^{-3}$ (bottom right).}
\label{fig:super2}
\end{figure}

In Fig.\ \ref{fig:super2}, we plot the temporal evolution of the
mean-square displacement for different ratios $d/l_{0}$ of channel width 
to length, the refractive index is $n_{l}=2$. For large ratios such as 
$d/l_{0} = 1$, superdiffusion is stronger since then photons
have more possibilities to realize very long effective steps. These
steps are visible in the insets that show the photon clouds at time
$t=10^{5} l_{0}/c$ for $d/l_{0} = 1$ and $10^{-3}$ 
(see also Fig.\ \ref{fig:super1}). The clouds are not
spheres as for conventional diffusion, instead, due to the long effective
steps, they exhibit peaks along the six equivalent directions of the 
honeycomb foam. The peaks are more pronounced for stronger superdiffusion.

\subsection{Diffusion in Disordered Voronoi Foams} \label{voronoi}

\begin{figure}
\includegraphics[width=0.9\columnwidth]{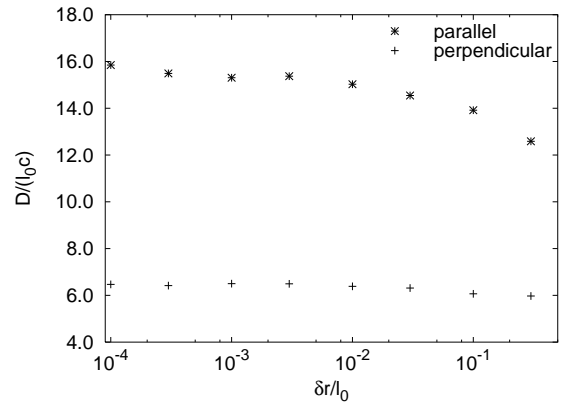}
\caption{Diffusion constant $D$ in units of $l_{0}c$ as a function of 
disorder $\delta r/l_{0}$ in the model foam with $d/l_{0}=0.1$, $n_{g}=1$ 
and $n_{l}=1.33$.}
\label{fig:diff1}
\end{figure}

As soon as we introduce disorder into the Voronoi foam, superdiffusion
vanishes and conventional diffusion occurs. This is not unexpected since 
disorder inhibits long effective steps of a photon as explained in the
previous section. In Fig.\ \ref{fig:diff1}
we plot the diffusion constant $D$, for both parallel and perpendicular
polarization of the light wave, as a function of $\delta r$, which 
quantifies disorder in our Voronoi foams. We expect the diffusion constant to 
diverge when $\delta r/l_{0}\rightarrow 0$ because of the superdiffusiv 
behaviour at $\delta r/l_{0}=0$. Our simulations indicate (note the 
logarithmic scale for $\delta r$) that the divergence is quite weak 
especially for the perpendicular polarization state. On the whole range of 
$\delta r$, a clear decrease of the diffusion constant with 
increasing disorder is visible. We understand this observation
since correlations between successive photon steps are destroyed more 
easily in a disordered foam.
Furthermore, the diffusion constant for parallel polarization is larger
compared to the perpendicular case due to its smaller reflectance $r_{\|}$,
which even becomes zero at the Brewster angle characterized by 
$\beta_{l} + \beta_{g} = \pi/2$.

\begin{figure}[b]
\includegraphics[width=0.9\columnwidth]{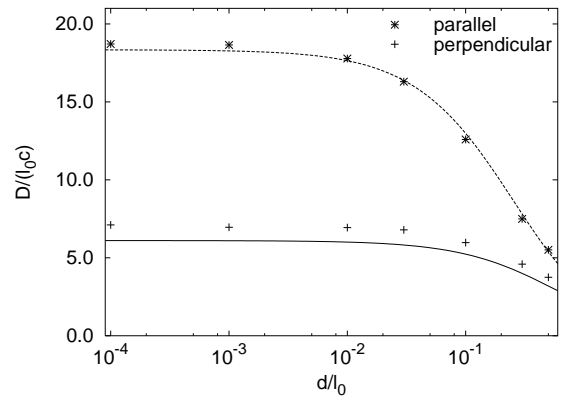}
\caption{Diffusion constant $D$ in units of $l_{0}c$ as a function of the 
channel width to length ratio $d/l_{0}$ in a disordered model foam with 
$\delta r/l_{0}=0.3$, $n_{g}=1$ and $n_{l}=1.33$. Symbols indicate numerical 
results, the lines follow from Eq.\ (\ref{eqn:D}) derived by a a 
one-dimensional analytical model.}
\label{fig:diff2}
\end{figure}

\begin{figure}
\includegraphics[width=0.9\columnwidth]{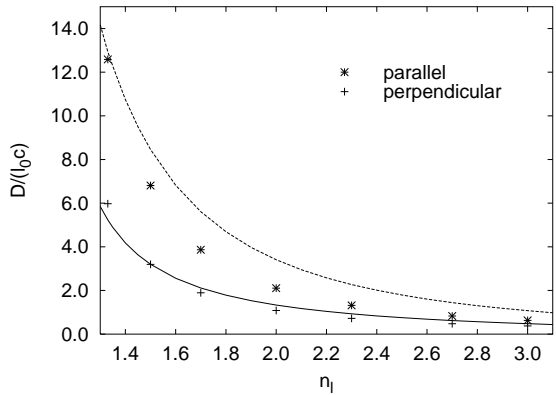}
\caption{Diffusion constant $D$ in units of $l_{0}c$ as a function of 
refractive index $n_{l}$ of the liquid phase in a disordered model 
foam with $\delta r/l_{0}=0.3$, $d/l_{0}=0.1$ and $n_{l}=1$. Symbols indicate 
numerical results, the lines follow from Eq.\ (\ref{eqn:D}) derived by a 
one-dimensional analytical model.}
\label{fig:diff3}
\end{figure}

In Fig.\ \ref{fig:diff2} the dependence of the diffusion constant on the 
channel width to length ratio $d/l_{0}$ is shown for the most disordered 
foam in our numerical treatment, i.e., for $\delta r/l_{0}=0.3$. The
ratio $d/l_{0}$ is directly connected to the liquid fraction $\varepsilon$ 
via Eq.\ (\ref{eqn:eps}) which offers an explanation for the observed
decrease of $D$ with increasing $d$. When the liquid fraction $\varepsilon$ 
becomes larger, the photons spend more time in the liquid phase where 
they move with a smaller velocity compared to the gaseous phase. Furthermore, 
more photons exhibit total reflection at the liquid-gas interface, i.e., 
photon channelling is more pronounced. Note also that for 
$d,\varepsilon \rightarrow 0$, the diffusion constant approaches
a finite value. This is in clear contrast to experiments\ \cite{durApp}
and we will comment on it in our conclusions. We have developed a 
one-dimensional random-walk model which includes a photon-channelling state
and whose details are explained in the following section. Its prediction
for the diffusion constant (see lines in Fig.\ \ref{fig:diff2}) gives
a remarkable quantitative agreement with the numeric results in 
Fig.\ \ref{fig:diff2}. Finally, Fig.\ \ref{fig:diff3} demonstrates
that the diffusion constant decreases with increasing $n_{l}$. This agrees
with intuition since the reflectance increases and the photons in the
liquid phase move slower. The random-walk model (see lines in 
Fig.\ \ref{fig:diff3}) again gives a good description for the observed
behavior with some quantitative deviations for the parallel polarization
state.

\subsection{A One-Dimensional Model for Photon Channelling}

\begin{figure*}
\includegraphics[width=1.7\columnwidth]{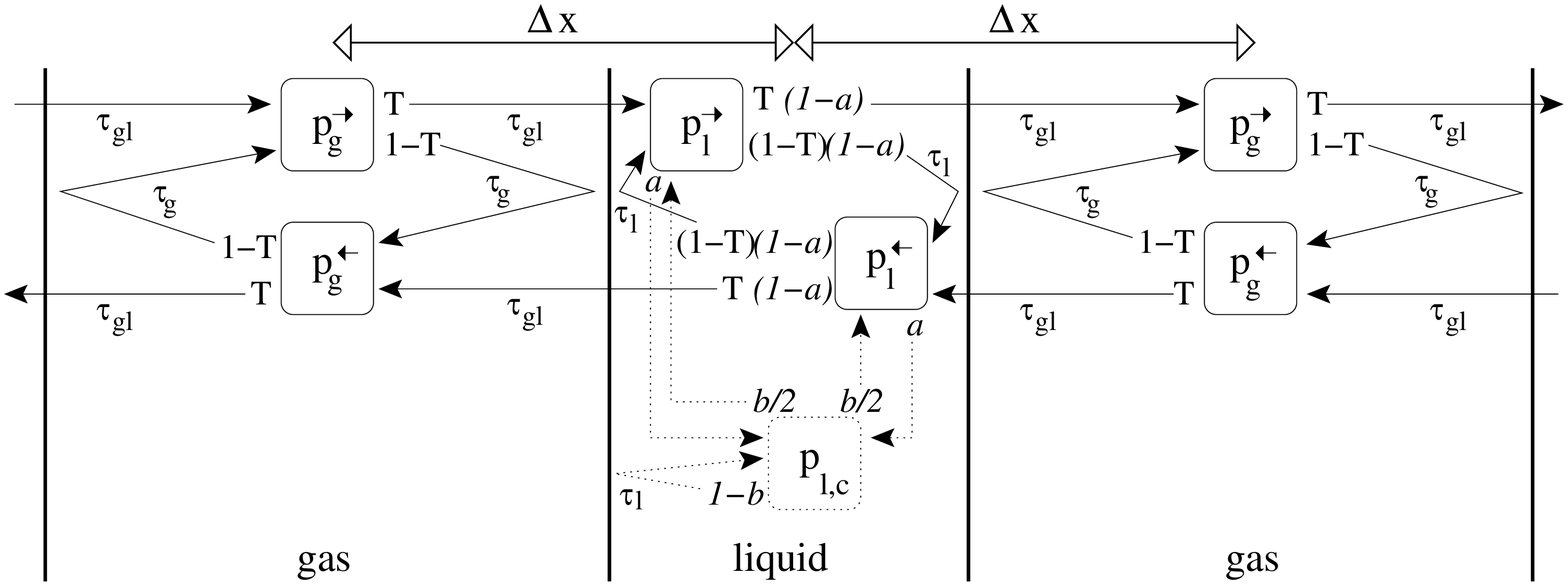}
\caption{One-dimensional random-walk model in a periodic sequence of
gaseous and liquid regions. Five photon states 
with probabilities $p_{g}^{\rightarrow}$, $p_{g}^{\leftarrow}$, 
$p_{l}^{\rightarrow}$, $p_{l}^{\leftarrow}$ and $p_{l,c}$ are introduced. 
The subscript stands for the phase, in which the photon resides, 
and the superscript arrow for its direction of motion. 
Photons with angles $\beta_{l}>\beta^{*}$, i.e., photons undergoing total 
reflection in the liquid phase, are in the photon channelling state 
with probability $p_{l,c}$. The possible transitions between the states 
are given by arrows. The transition probabilities and times are explained 
in the text. Note that the transitions from and into the photon 
channelling state are instantaneous.}
\label{fig:model}
\end{figure*}

To explain the main features of photon channelling in disordered foams,
we have developed a one-dimensional random-walk model. We consider 
a sequence of gaseous and liquid regions with a lattice constant $\Delta x$, 
as illustrated in Fig.\ \ref{fig:model}. On a length scale larger than
$\Delta x$, we then introduce the coarse-grained probabilities 
$p_{g}^{\rightarrow}(x,t)$, $p_{g}^{\leftarrow}(x,t)$, 
$p_{l}^{\rightarrow}(x,t)$ and $p_{l}^{\leftarrow}(x,t)$ for a photon to 
be at time $t$ at the position $x$ and in a state specified by the indices.
Here the lower index means that the photon resides in the gaseous ($g$) or
liquid ($l$) phase and the arrow of the upper index gives the direction of
motion. All photons are transmitted through the liquid-gas interface with
a probability $T$ or they are reflected with a probability $1-T$. 
In Fig.\ \ref{fig:model} all possible transitions between the four states 
are indicated by solid lines, the transmission probabilities are given in 
non-italic writing. Reflected photons return to their original position,
which for concreteness we locate in the middle of each region, after a 
time $\tau_{g}$ or $\tau_{l}$. Transmitted photons move from the gaseous to
the liquid phase or vice versa during the time 
$\tau_{gl}=(\tau_{g}+\tau_{l})/2$.

So far, all photons can traverse the liquid-gas interface with a probability
$T$. To allow for photon channelling in our model, i.e., photons which
exhibit total reflection, we introduce a fifth state located in the 
liquid phase with the probability $p_{l,c}(x,t)$. Within the liquid phase,
photons travelling to the right or left switch instantaneously into 
the photon channelling state with a probability $a$. They return to one
of the ``moving'' states with a probability $b/2$. The numbers $a$ and $b$
are related to each other as we will demonstrate below. With probability 
$1-b$, a photon in the photon channelling state is reflected at the 
liquid-gas interface and returns to the same state after the time 
$\tau_{l}$. All possible transitions connected to photon channelling are
indicated by dashed lines in Fig.\ \ref{fig:model}.

The following master equations now quantify all transitions between 
different photon states:
\begin{widetext}
\begin{eqnarray}
\nonumber
p_{g}^{\rightarrow}(x,t)&=&(1-T)p_{g}^{\leftarrow}
(x,t-\tau_{g})+T(1-a)p_{l}^{\rightarrow}(x-\Delta x,t-\tau_{gl})\\
\nonumber
p_{g}^{\leftarrow}(x,t)&=&(1-T)p_{g}^{\rightarrow}(x,t-\tau_{g})+
T(1-a)p_{l}^{\leftarrow}(x+\Delta x,t-\tau_{gl})\\
\label{mastereq}
p_{l}^{\rightarrow}(x,t)&=&Tp_{g}^{\rightarrow}(x-\Delta x,t-\tau_{gl})+
(1-T)(1-a)p_{l}^{\leftarrow}(x,t-\tau_{l})+\frac{1}{2}bp_{l,c}(x,t)\\
\nonumber
p_{l}^{\leftarrow}(x,t)&=&Tp_{g}^{\leftarrow}(x+\Delta x,t-\tau_{gl})+
(1-T)(1-a)p_{l}^{\rightarrow}(x,t-\tau_{l})+\frac{1}{2}bp_{l,c}(x,t)\\
\nonumber
p_{l,c}(x,t)&=&ap_{l}^{\rightarrow}(x,t)+ap_{l}^{\leftarrow}(x,t)+(1-b)
p_{l,c}(x,t-\tau_{l}).
\end{eqnarray}
We are interested in the diffusive behavior of the total
probability $p(x,t) = p_{g}^{\rightarrow} + p_{g}^{\leftarrow} +
p_{l}^{\rightarrow} + p_{l}^{\leftarrow} + p_{l,c}$ that should be visible on
large length and time scales. We therefore Fourier transform the master
equations\ (\ref{mastereq})
and then Taylor expand all the coefficients up to first order in 
the frequency $\omega$ and up to second order in the wave number $k$. Thus
we obtain a set of linear equations:
\begin{equation}
\left(
\begin{array}{ccccc}
-1 & (1-T)f(\tau_{g}) & T(1-a)g^{-} & 0 & 0\\
(1-T)f(\tau_{g}) & -1 & 0 & T(1-a)g^{+} & 0\\
Tg^{-} & 0 & -1 & (1-T)(1-a)f(\tau_{l}) & b/2\\
0 & Tg^{+} & (1-T)(1-a)f(\tau_{l}) & -1 & b/2\\
0 & 0 & a & a & -1+(1-b)f(\tau_{l})\\
\end{array}
\right)
\left(
\begin{array}{c}
p_{g}^{\rightarrow}(\omega,k)\\
p_{g}^{\leftarrow}(\omega,k)\\
p_{l}^{\rightarrow}(\omega,k)\\
p_{l}^{\leftarrow}(\omega,k)\\
p_{l,c}(\omega,k)
\end{array}
\right)=0
\label{13}
\end{equation}
\end{widetext}
where $f(\tau)=1-i\omega \tau$ and $g^{\pm}=1-i\omega\tau_{gl}\pm ik\Delta 
x\pm \omega k \tau_{gl} \Delta x-k^{2}\Delta x^{2}/2$. Now, each 
probability in Eqs.\ (\ref{13}) obeys the same relation, e.g.,
$\mathrm{det}\mathbf{M} \,\, p_{g}^{\rightarrow}(\omega,k) = 0$ where 
$\mathrm{det}\mathbf{M}$ denotes the determinant of the coefficient
matrix in (\ref{13}). To leading order in $\omega$ and $k$, it reads 
$\mathrm{det}\mathbf{M} \propto i\omega +Dk^{2}$. Thus, our model indeed 
reproduces diffusive behavior and gives a formula for the diffusion constant:
\begin{equation}
\label{eqn:D}
D=\frac{2\Delta x^{2}bT(1-a)(2-a)}{\left[\tau_{g}b(1-a)+\tau_{l}(a+b-2ab)
\right]\left[3aT-2a-4T+4\right]}
\end{equation}
(note that $\tau_{gl}=(\tau_{g}+\tau_{l})/2$ was already used).

We now make contact with our two-dimensional model foams. Photons in the 
four states $p_{g}^{\rightarrow}$, $p_{g}^{\leftarrow}$, 
$p_{l}^{\rightarrow}$ and $p_{l}^{\leftarrow}$ correspond or lead to
photons in the liquid phase whose angle with respect to the interface normal 
obeys $\beta < \beta^{*}$, i.e., they do not exhibit total reflection.
Thus we choose $T=T_{g\rightarrow l}$ where $T_{g\rightarrow l}$ 
is the average transmission probability introduced in section\ 
\ref{channel.analytic}. 

In stationnary equilibrium, the flux of photons
into and out of the photon channnelling state has to be equal:
$ap_{l}^{\rightarrow}+ap_{l}^{\leftarrow}=bp_{l,c}$. This implies a relation
between $a$ and $b$:
\begin{equation}
\frac{a}{b}=\frac{p_{l,c}}{p_{l}^{\rightarrow}+p_{l}^{\leftarrow}}=
\frac{\int_{\beta^{*}}^{\pi/2}p(\beta_{l})d\beta_{l}}{
\int_{0}^{\beta^{*}}p(\beta_{l})d\beta_{l}}=\frac{n_{l}}{n_{g}}-1 \enspace.
\end{equation}
Here we have used the formalism of section\ \ref{channel.analytic} to
relate $p_{l,c}$ and $p_{l}^{\rightarrow}+p_{l}^{\leftarrow}$ to the 
probability $p(\beta_{l})$ and then have assumed an equal angular 
distribution of the photons, thus $p(\beta_{l})\propto \cos(\beta_{l})$, 
as in section \ref{channel.analytic}. The quantity $b$ denotes the 
probability for a photon to leave the photon channelling state. In our
two-dimensional model foams this can only occur at a junction where
three channels meet since in the channel itself reflections do not change 
$\beta>\beta^{*}$. The number of reflections necessary to reach a junction
scales as $d^{-1}$, we therefore assume $b\propto d/l_{0}$. Since the
constant of proportionality is of the order of one, we take $b=d/l_{0}$
to compare the one-dimensional model to the simulation data. 
In Fig.\ \ref{fig:diff2}, e.g., variations in $b$ will shift the region where
the diffusion constant decreases.

\begin{figure}
\includegraphics[width=0.7\columnwidth]{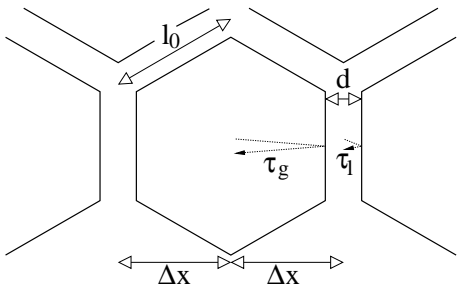}
\caption{The parameters $\Delta x$, $\tau_{g}$ and $\tau_{l}$
of the one-dimensional photon-channelling model in relation
to a two-dimensional model foam: $\Delta x=\sqrt{3}l_{0}/2$, 
$\tau_{g}=n_{g}(\sqrt{3}l_{0}-d)/c$ and $\tau_{l}=n_{l}d/c$.}
\label{fig:sechseck}
\end{figure}

The lattice constant $\Delta x$ determines the diffusion constant in the
limit where the liquid region tends to zero ($d \rightarrow 0$). From 
Eq. (\ref{eqn:D}) one finds 
\begin{equation}
D(d \rightarrow 0) =  \frac{1}{2} \frac{T}{1-T} \Delta x \, c.
\end{equation}
In our model foams, $\Delta x$ corresponds to half of the typical bubble 
size. To compare the diffusion constant of Eq.\ (\ref{eqn:D}) to our 
simulation results, we choose $\Delta x$ and the times $\tau_{l}$ and 
$\tau_{g}$ as shown in Fig.\ \ref{fig:sechseck}, i.e., 
$\Delta x=\sqrt{3}l_{0}/2$, $\tau_{g}=n_{g}(\sqrt{3}l_{0}-d)/c$ and 
$\tau_{l}=n_{l}d/c$. As illustrated in Fig.\ \ref{fig:diff2} and 
\ref{fig:diff3}, this gives an excellent agreement with our simulation
data. So the one-dimensional model explains the main features of our
approach to photon channelling very well. Absolutely crucial for the
success of the model is the introduction of the photon-channelling state.

\section{Conclusions}\label{conclus}

Photon channelling found in experiments by Gittings {\em et al.}
is an interesting concept which increases our understanding of
the diffusive transport of light in foams. To contribute to its
further study, we have simulated the transport of light in 
two-dimensional Voronoi foams based on ray optics. In perfect honeycomb 
structures, we find superdiffusive behavior which is connected to
the motion of photons in the liquid channels. However, as soon as
we introduce disorder into the foams, light transport becomes diffusive.
This agrees with our finding that the diffusion constant $D$ only exhibits
a weak divergence for decreasing disorder parameter $\delta r$. 
Furthermore, according to our simulations and in agreement with 
intuition, $D$ decreases with increasing refractive index $n_{l}$ and 
also with increasing channel-width-to-length ratio $d/l_{0}$. 
We are able to model this dependence with the help of a one-dimensional 
random walk model but only when we introduce a photon-channelling state. 
Therefore, the importance of photon-channelling becomes evident.
The decrease of $D$ with increasing $d/l_{0}$ or liquid fraction 
$\varepsilon$ is in qualitative agreement with measurements by
Vera {\em et al.\/}\ \cite{durApp}. However, there is an important
difference; whereas Vera {\em et al.\/} observe a divergence of the
diffusion constant for $\varepsilon \rightarrow 0$, it assumes a
finite value in our simulations and analytic model. The reason is that
we consider independent transmission and reflection events at each single 
liquid-gas interface. If the film thickness becomes sufficiently thin, the
effective reflectance and transmittance of thin films seems to be
more appropriate, where interference effects are taken into account\ 
\cite{Miri04}. They lead to a strong decrease of the reflectance 
for $\varepsilon \rightarrow 0$ and therefore to a divergence of $D$.

Building on the work of Gittings {\em et al.}\ \cite{Gittings2004}, 
we are able to justify their observed relation between the photon's 
path-length fraction $f$ in the liquid phase and the liquid fraction 
$\varepsilon$ without any free parameters. Their further observation
that photon channelling only occurs in the region 
$0.04 < \varepsilon < 0.2$ will help to improve our understanding of
light propagation in foams. For $\varepsilon < 0.04$, the extension
of the Plateau borders and films are probably too small that pure
ray optics is applicable and therefore photon channelling breaks down.
In this parameter region, we suggest that the reflectance and 
transmittance of thin films which include interference effects, as already 
mentioned in the last paragraph, are the appropriate quantities to
model the photons' random walk. However, we have no understanding
why photon channelling ceases for $\varepsilon > 0.2$. Especially, our
analytic considerations in Sec.\ \ref{channel.analytic} employ very 
general arguments without relying on a concrete foam structure, so it 
is not evident why they should break down for larger volume fractions.

Nevertheless, we show in this article that photon channelling
in liquid foams can be understood on the basis of ray optics
through analytic considerations and simulations using simple
model foams. It seems that details of the foam structure are
unimportant. We are confident that the results presented in this article add another piece to the puzzle of explaining diffusive light transport in cellular structures such as aqueous foams.

\begin{acknowledgments}
We would like to thank D. Durian, R. H\"{o}hler, R. Maynard, G. Maret 
and S.E. Skipetrov for helpful discussions, and J.~R. Shewchuk for
making the program {\em Triangle\/} publicly available. 
H.S. acknowledges financial support from 
the Deutsche Forschungsgemeinschaft under Grant No. Sta 352/5-2.
MF.M and M.S. thank the International Graduate College at the 
University of Konstanz for financial support.
\end{acknowledgments}



\end{document}